\documentclass[a4paper,fleqn,usenatbib]{mnras}

\usepackage{newtxtext,newtxmath}

\usepackage[T1]{fontenc}
\usepackage{ae,aecompl}

\usepackage{graphicx}	
\usepackage{amsmath}	
\usepackage{amssymb}	

\title[Coherent radio emission from HD\,133880]{Discovery of electron cyclotron MASER emission from the magnetic Bp star HD\,133880 with the Giant Metrewave Radio Telescope}

\author[B. Das et al.]{
Barnali Das,$^{1}$\thanks{E-mail: barnali@ncra.tifr.res.in}
Poonam Chandra,$^{1,2}$
and Gregg A. Wade$^{3}$
\\
$^{1}$National Centre for Radio Astrophysics, Tata Institute of Fundamental Research,  Pune University Campus, Pune-411007, India\\
$^{2}$Department of Astronomy, Stockholm University, AlbaNova, SE-106 91 Stockholm, Sweden\\
$^{3}$Department of Physics, Royal Military College of Canada, PO Box 17000, Station Forces, Kingston, ON K7K 7B4, Canada
}

\date{Accepted 2017 November 24. Received 2017 November 17; in original form 2017 September 24}

\pubyear{2017}

\begin{document}
\label{firstpage}
\pagerange{\pageref{firstpage}--\pageref{lastpage}}
\maketitle

\begin{abstract}
We report the discovery of  coherent radio emission from the young, rapidly-rotating magnetic Bp star  HD\,133880 at a frequency of 610\,MHz with the Giant Metrewave Radio Telescope (GMRT). This is only the second magnetic star in which coherent radio emission has been detected. 
 In our observations of HD\,133880 covering the full rotational cycle of the star (except for a phase window $0.17-0.24$), 
 we witness an abrupt
order-of-magnitude flux enhancement along with $\approx100$ percent right circular polarization.  We attribute this phenomenon to coherent  Electron Cyclotron MASER Emission. We attribute the lack of left circularly polarised emission to the asymmetric topology of the star's magnetic field.
The phase of enhancement, $0.73$, differs from the previously reported phase of enhancement, $0.16$, \citep[at 610 MHz][]{chandra15}
by one-half cycle. However, no flux enhancement is found at phase $0.16$ in our data, which could be due to an unstable or drifting emission region, or a consequence of the reported changes of the star's rotational period. Either of these factors could have shifted the enhancement to the above-mentioned phase window not sampled by our observations.
\end{abstract}

\begin{keywords}
stars: individual: HD\,133880 -- stars: magnetic field -- masers -- polarization
\end{keywords}



\section{Introduction}  \label{sec:intro}

Magnetic hot stars are O- and B-type stars characterised by the presence of strong, organized mostly dipolar \citep{babcock49} magnetic fields in their photospheres.
Initially, models for radio emission in magnetic stars  were proposed by \citet{drake87,andre88,linsky92}.
These ideas were unified and given quantitative form by \citet{trigilio04}. 
According to this model, the radio emission is gyro-synchrotron in nature, as a result of the interaction of the 
magnetic field with a radiatively-driven wind. 
Between the ``inner magnetosphere'' dominated by the magnetic field and  the ``outer magnetosphere'' dominated by the wind kinetic pressure, 
magnetic field lines are opened up by the wind, forming an equatorial current 
sheet in which electrons are accelerated to relativistic energies. Eventually, these electrons return towards the stellar surface following the field lines and 
emit radio photons by the gyro-synchrotron process, 
 at a frequency close to the harmonics of the local gyro-frequency, which is a function of 
the local magnetic field strength. 
Because the magnetic field decreases in strength with distance from the star, emission at higher frequencies is generated closer to the star and vice-versa. 
As a consequence, the study of radio emission from magnetic stars at different frequencies allows one to probe the circumstellar environment at different distances from the star.

Evidence supporting a gyro-synchrotron origin of the radio emission from magnetic stars includes the observed correlation between the variability of radio light-curves with that of the line-of-sight magnetic 
field and rotation period of the star, the high brightness temperature,  and the absence of a high degree of circular polarization \citep{drake87,linsky92,leto06}. 
 However, in the rapidly-rotating magnetic Bp star CU Vir, \citet{trigilio00} reported the detection 
 of highly directive, 100 per cent right circularly polarized  coherent emission at 1.4\,GHz..
The authors proposed Electron Cyclotron MASER Emission (ECME) as the most likely mechanism \citep{trigilio00,trigilio08}.
This emission is explainable in the light of the 3D emission model by \citet{leto16}. The highly energetic electrons in the current sheet, while coming back 
towards the star following the converging field lines, give rise to the condition for magnetic mirroring. 
The electrons with sufficiently small pitch angles are lost from the region creating a loss-cone anisotropy, and if the plasma frequency 
is much smaller than the local gyro-frequency,
masing action can occur, which is more favourable at lower frequencies  \citep{melrose82}. ECME is favoured at lower frequencies, since, as a direct consequence of the resonance conditions, the number of resonant electrons decreases towards higher frequencies (Eq. 3a and 3b, \citet{melrose82})  
However, 
if  the corresponding gyro-frequency becomes less than the local plasma frequency, plasma emission
should replace ECME as the dominant coherent mechanism at low enough frequencies. 
One key difference between the two mechanisms is the role of the magnetic field. 
In case of ECME, the magnetic field 
determines the frequency and the direction of emission of the radiation, whereas the role of magnetic field is negligible in plasma emission.

\par
In this paper, we report the discovery of coherent emission from the magnetic star HD\,133880 (= HR\,Lup= HR\,5624) at 610\,MHz with the Giant Metrewave Radio
Telescope (GMRT).  HD\,133880 is a  young, Bp type star with a rotation period of $\approx 0.88$~d. \citep{waelkens85,landstreet90,bailey12,kochukhov17}.
It has an unusual magnetic topology, as the rotation-phased longitudinal (line-of-sight) magnetic field ($B_z$) differs significantly from the typical sinusoidal variation.
This led \citet{landstreet90} to propose that the star's magnetic field is a combination of a dipole component and a much stronger quadrupole component.  \citet{bailey12} reported the 
strength of dipolar and quadrupolar components as $B_d=-9600 \pm 1000$ G and $B_q=-23200 \pm 1000 $ G,  respectively. 
However, this was contradicted by \citet{kochukhov17}, who demonstrated that the phase-resolved Stokes $V$ profiles of HD\, 133880 are best reproduced using a weaker, distorted dipole, topology. 
In their study, they found that the star has a weak but extended positive field region and a stronger but smaller negative field region. They also found the mean magnetic field strength to be much smaller ($\approx 4$ kG) than that ($\approx 11-23$ kG) reported by \citet{bailey12}.
\par 

The period of the star too has been modified many times. \citet{waelkens85} first estimated the rotation period as 0.877459 days which was revised  to 0.877485 days
 \citep{landstreet90}. \citet{bailey12} again revised this period to 0.877476 days in order to remove the offset between their measurements of magnetic field and those of \citet{landstreet90}. 
 However,  recently, \citet{kochukhov17} reported the period to be 0.877483 days - close to the value proposed by \citet{landstreet90}. The reported values of the periods differ significantly from one another, and can only be reconciled if the period is assumed to vary with time \citep{kochukhov17}.

 \par
In this paper, we adopt the most recent period of 0.877483 days \citep{kochukhov17} and the reference Julian date 2445472.000 \citep{bailey12} for all calculations.
 Hence we use the ephemeris $$ JD=2445472.000+(0.877483)\cdot E. $$

 \par
       
Unlike CU Vir, which has been studied extensively in radio bands, 
\citep{trigilio00,trigilio04,leto06,trigilio08,ravi10,trigilio11,lo12}, HD\,133880 has been studied very little. 
In this Letter, we present observations of HD\,133880 at 610 MHz covering nearly full rotation period which enable us to see the coherent nature of the radio emission.
This paper is structured as follows: We describe our observations and data analysis in \S  \ref{sec:data}. Our  results  are
presented in \S \ref{sec:results}  and finally we  discuss and summarise our main results in \S \ref{sec:summary}.

\section{Observations and data reduction}\label{sec:data}

We observed HD\,133880 at a frequency of 610\,MHz with the GMRT
for seven consecutive nights, from 2016 February 5 until 2016 February 11 in two polarizations, right circular  $RR^*$ (Stokes $(I+V)$) and 
left circular  $LL^*$ (Stokes $(I-V)$). The observations were carried out for  $4$ hours each day at the same Indian Standard Time, so as to achieve a full 
rotational phase coverage.  
Including overheads due to the flux calibrator (3C286)  and 
phase calibrator (J1626$-$298), the final on-source time
at each epoch varied from $\sim 2.5$ hours to 3 hours. 
The flux calibrator was also used as bandpass calibrator.  The phase calibrator was observed after every target (i.e. HD\,133880) scan of 35~minutes.
The total bandwidth of the observations was 32\,MHz, divided equally into 256 channels. 
However, for the final imaging, only the central $\approx 25$ MHz were used. Initial flagging and calibration were done using the Common Astronomy Software Applications \citep[CASA, ][]{mcmullin07}
package, and  the imaging and subsequent self-calibrations were done using the Astronomical Image Processing System (AIPS).
For each data set, an average sky map was produced using the standard procedures in AIPS. All non-target sources were then subtracted from the visibility data using the AIPS  tasks `BOX2CC', `CCEDT' and `UVSUB'. 
The resulting single source visibility data were then used to get an image for every scan, independently for  the $RR^*$ and $LL^*$ polarizations.
The flux density values were obtained using the AIPS task JMFIT. 

\par
This method has one drawback. 
During self calibration, it is assumed that the sky is stationary and the only quantity which can vary with time is the antenna gain. 
Because of this, the actual temporal variation of a source is likely to be suppressed. 
However, since the 
self-calibration is dominated by much brighter sources  (the brightest ones are $\sim 100$ mJy) in the field of view (FoV), 
the variation of the flux with time can still be recovered \citep{pr84}. 
Thus it is safe to apply this method to find the base flux of the star which is much smaller \citep[$\sim 1$ mJy, ][]{chandra15} and roughly constant.
The strategy that we followed was to employ the above method if the average flux at each epoch was of the order of the base flux. 
In our data, out of seven data sets, six were found to fulfil this condition. The data set obtained on 2016 February 08 did not comply with this condition and thus
each scan was imaged independently.

\par
To demonstrate the robustness of this method, we estimated the flux density of a test source SUMSS J150757-403711 (RA: $15^h07^m57^s.011$, Dec: $-40^o37'13.65''$) in the FoV at each epoch.  We did not find any significant variation  over various epochs as well as in the right and left circular
polarizations. The maximum variation of 
$\sim 10$ per cent  is consistent with the known systematic errors of GMRT data at 610\,MHz \citep{ck17}.

\section{Results }
\label{sec:results}

In Table \ref{tab:LLRR_data}, we have listed the details of the observations and flux densities in the $RR^*$ and $LL^*$ polarizations.
The errors in the flux values correspond to fitting errors obtained from `JMFIT'. Except for the data taken on 2016 February 5, the mean HJD corresponds
 to the mean epoch of the data averaged for each scan, with uncertainties reflecting the 
  length of the time intervals over which averaging was performed.
The data for 2016 February 5 were found to be too noisy; as a result, we averaged over a longer time.

As is evident from Table \ref{tab:LLRR_data}, there are flux enhancements between phases 0.70 and 0.77 in the 2016 February 8 data with a peak at a rotational phase of 0.731 in right 
circular polarization $RR^*$.  We further divided the scans corresponding  the enhanced emission into three  $\sim 10$ minutes
subscans. Finally, the subscan at which the flux attained its maximum was further divided into five segments each of two minutes duration.
The resulting values are reported in Table \ref{tab:RR_data_8feb}.

\begin{table}
{\tiny
\caption{Variation of the flux density of HD\,133880 with its rotational phase} \label{tab:LLRR_data}
\begin{tabular}{cccc|c}
\hline
  Date      &.       Mean HJD              &      Phase       &  $LL^*$ Flux  & $RR^*$ Flux   \\ 
        & & & mJy  & mJy \\
        \hline
05-Feb-2016 & 2457423.56148$\pm$0.028 & 0.277$\pm$0.032 & 2.7$\pm$0.2  & 1.2$\pm$0.2  \\
            & 2457423.63349$\pm$0.037 & 0.359$\pm$0.042 & 1.0$\pm$0.2  & 1.2$\pm$0.2 \\
06-Feb-2016 & 2457424.51880$\pm$0.012 & 0.368$\pm$0.014 & 1.2$\pm$0.2  & 1.7$\pm$0.2 \\
            & 2457424.55031$\pm$0.012 & 0.404$\pm$0.014 & 1.5$\pm$0.2  & 1.8$\pm$0.2 \\
            & 2457424.58162$\pm$0.012 & 0.440$\pm$0.014 & 1.4$\pm$0.2  & 1.6$\pm$0.2 \\
            & 2457424.61621$\pm$0.012 & 0.479$\pm$0.014 & 1.3$\pm$0.2  & 2.0$\pm$0.2 \\
07-Feb-2016 & 2457425.52611$\pm$0.012 & 0.516$\pm$0.014 & 1.3$\pm$0.2  & 2.5$\pm$0.2  \\
	        & 2457425.55724$\pm$0.012 & 0.551$\pm$0.014 & 1.3$\pm$0.3  & 2.0$\pm$0.3  \\
	        & 2457425.58856$\pm$0.012 & 0.587$\pm$0.014 & 1.3$\pm$0.4  & 2.0$\pm$0.4  \\
	        & 2457425.61951$\pm$0.012 & 0.622$\pm$0.014 & 1.6$\pm$0.2 & 1.6$\pm$0.2  \\   
08-Feb-2016 & 2457426.49958$\pm$0.012 & 0.625$\pm$0.014 & 1.1$\pm$0.2 & 2.3$\pm$0.2 \\
            & 2457426.53053$\pm$0.012 & 0.660$\pm$0.014 & 1.4$\pm$0.2  & 2.0$\pm$0.2 \\
            & 2457426.56165$\pm$0.012 & 0.696$\pm$0.014 & 3.0$\pm$0.2  & 7.9$\pm$0.2 \\
            & 2457426.59261$\pm$0.012 & 0.731$\pm$0.014 & 1.3$\pm$0.2  &16.8$\pm$0.2  \\
            & 2457426.62374$\pm$0.012 & 0.767$\pm$0.014 & 0.9$\pm$0.2  & 2.5$\pm$0.2 \\      
09-Feb-2016 & 2457427.51724$\pm$0.012 & 0.785$\pm$0.014 & 1.2$\pm$0.2  & 1.4$\pm$0.2 \\
            & 2457427.54819$\pm$0.012 & 0.820$\pm$0.014 & 1.2$\pm$0.2  & 1.2$\pm$0.2 \\
            & 2457427.57933$\pm$0.012 & 0.856$\pm$0.014 & 1.6$\pm$0.2 & 1.2$\pm$0.2 \\
            & 2457427.61381$\pm$0.012 & 0.895$\pm$0.014 & 1.6$\pm$0.2  & 1.2$\pm$0.2 \\
10-Feb-2016 & 2457428.51607$\pm$0.012 & 0.923$\pm$0.014 & 1.8$\pm$0.2  & 1.7$\pm$0.2 \\
            & 2457428.54701$\pm$0.012 & 0.958$\pm$0.014 & 1.6$\pm$0.2  & 1.4$\pm$0.2 \\
            & 2457428.57796$\pm$0.012 & 0.994$\pm$0.014 & 1.6$\pm$0.2  & 1.7$\pm$0.2 \\
            & 2457428.60910$\pm$0.012 & 0.029$\pm$0.014 & 1.4$\pm$0.2  & 1.2$\pm$0.2 \\
            & 2457428.63697$\pm$0.012 & 0.061$\pm$0.014 & 1.1$\pm$0.2  & 1.6$\pm$0.2 \\
11-Feb-2016 & 2457429.50501$\pm$0.012 & 0.050$\pm$0.014 & 2.1$\pm$0.2  & 1.8$\pm$0.2 \\
            & 2457429.53596$\pm$0.012 & 0.086$\pm$0.014 & 1.7$\pm$0.2  & 1.9$\pm$0.2 \\
            & 2457429.56690$\pm$0.012 & 0.121$\pm$0.014 & 1.6$\pm$0.2  & 2.1$\pm$0.2 \\
            & 2457429.59785$\pm$0.012 & 0.156$\pm$0.014 & 1.9$\pm$0.2  & 1.8$\pm$0.2  \\
\hline
\multicolumn{5}{l}{$^a$ $LL^*$ polarization is Stokes $(I-V)$ and $RR^*$ polarization is Stokes $(I+V)$. }
\end{tabular}
}
\end{table}

\begin{table}
\begin{center}
{\scriptsize
\caption{Observed enhancements of the $RR^*$ polarization flux of HD\,133880 on 2016 February 8} \label{tab:RR_data_8feb}
\begin{tabular}{rrcc}
\hline
Mean HJD & Phase & Flux density  \\
 & & mJy  \\
\hline
2457426.55278$\pm$0.003 & 0.686$\pm$0.003 & 4.4$\pm$0.3  \\
2457426.56015$\pm$0.003 & 0.694$\pm$0.003 & 8.4$\pm$0.3  \\
2457426.56860$\pm$0.005 & 0.704$\pm$0.006 & 10.8$\pm$0.3 \\
2457426.58414$\pm$0.003 & 0.722$\pm$0.003 & 15.2$\pm$0.3  \\
2457426.58846$\pm$0.0007 & 0.7266$\pm$0.0008 & 20.4$\pm$0.8  \\
2457426.58984$\pm$0.0007 & 0.7281$\pm$0.0008 & 20.3$\pm$0.9  \\
2457426.59122$\pm$0.0007 & 0.7297$\pm$0.0008 & 21.5$\pm$0.9  \\
2457426.59262$\pm$0.0007 & 0.7313$\pm$0.0008 & 22.7$\pm$0.9  \\
2457426.59400$\pm$0.0007 & 0.7329$\pm$0.0008 & 22.3$\pm$0.9  \\
2457426.59955$\pm$0.005 & 0.739$\pm$0.006 & 14.4$\pm$0.3  \\
2457426.61528$\pm$0.003 & 0.757$\pm$0.003 & 5.4$\pm$0.3  \\
2457426.62722$\pm$0.008 & 0.771$\pm$0.009 & 1.0$\pm$0.2 \\             
\hline
  \end{tabular}
  }
  \end{center}
 \end{table}


\begin{table}
{\tiny
\caption{Stokes $V/I$ flux for HD\,133880 and the test source on 8 February 2016} \label{tab:V/I_data}
\begin{tabular}{cccc}
\hline
Mean HJD & Phase & Stokes $V/I$ & Stokes $V/I$\\
& & HD\,133880 & test source\\
\hline
2457426.49958$\pm$0.012 & 0.625$\pm$0.014 & 0.35$\pm$0.11 & 0.012$\pm$0.005 \\
2457426.53053$\pm$0.012 & 0.665$\pm$0.014 & 0.18$\pm$0.10 & 0.007$\pm$0.005 \\
2457426.56165$\pm$0.012 & 0.696$\pm$0.014 & 0.45$\pm$0.04 & -0.002$\pm$0.005 \\
2457426.59261$\pm$0.012 & 0.731$\pm$0.014 & 0.86$\pm$0.03 & 0.002$\pm$0.005 \\
2457426.62374$\pm$0.012 & 0.767$\pm$0.014 & 0.47$\pm$0.12 & 0.011$\pm$0.005 \\
\hline
  \end{tabular}
  }
 \end{table}

No significant  flux enhancement was obtained at any phase in left circular $LL^*$ polarization.
In order to ensure that the observed right circular polarization was indeed coming from the star and not due to any instrumental effect, 
we checked the polarization fraction (Stokes $V/I$) for 
the test source, SUMSS\,J150757-403711. Table \ref{tab:V/I_data} shows the polarization fraction for both sources.
It was observed that the maximum polarization that can result due to effects not intrinsic to the star is less than 2 per cent, 
consistent with the fact that the error in Stokes $V$ flux due to polarization leakage  in GMRT data is less than 5 per cent \citep{roy10}.
The maximum circular polarization measured in our observations of HD\,133880 is $\sim 90 \%$. However, if we consider the fact that the basal flux and the enhanced flux originate by different emission mechanisms, we can potentially remove the contribution of the basal flux to the total intensity while calculating the circular polarisation fraction in the peak emission. This involves the assumption that the circular polarisation fraction in the basal flux is negligible. In general, this assumption does not hold for the basal gyro-synchrotron emission from a magnetic star. In our case, we are rescued from this problem by the fact that the variation of circular polarisation due to gyro-synchrotron follows the longitudinal magnetic field ($B_z$) \citep[Figure 2 of][]{lim96} and we observed the enhancement when $B_z \approx 0$ (Figure \ref{fig:f1}), thus validating our assumption that the basal flux at the phase of enhancement is nearly unpolarised. Once we incorporate this, the maximum polarization fraction increases to $\approx 99 \%$.

\section{Discussion and Conclusions}
\label{sec:summary}

\begin{figure}
\centering
\includegraphics*[width=0.4\textwidth]{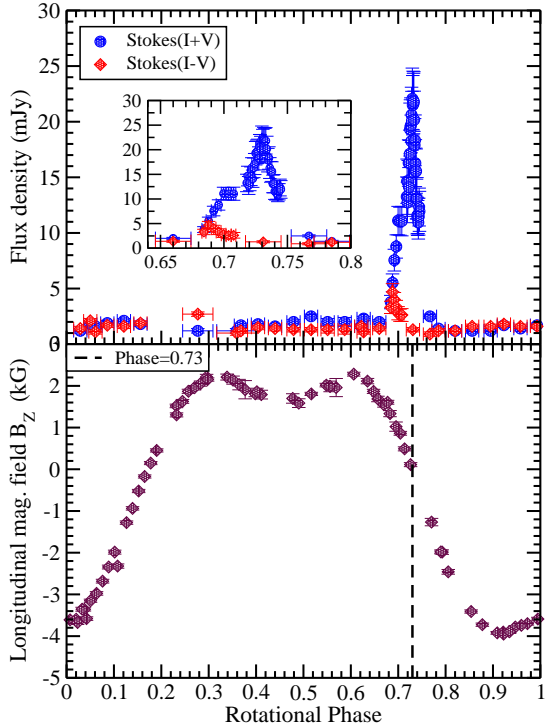}
\caption{Variation of the 610 MHz flux of HD 133880 (upper panel) and that of the line-of-sight magnetic field \citep[lower panel,][]{kochukhov17}  with rotational phase. 
The vertical dotted line on the right panel corresponds to the phase of enhancement. Error bars correspond to two $\sigma$. 
The inset in the upper panel shows the zoomed region where the $RR^*$ polarization flux is enhanced. \label{fig:f1}}
\end{figure}

The flux density variation for both polarizations $RR^*$ and $LL^*$  are shown in Figure \ref{fig:f1}.  We observe a significant flux enhancement ($\approx 23 $ mJy) in $RR^*$ polarization  at
phase $0.731\pm0.001$. To have an idea about the corresponding brightness temperature, we put the size of the emitting region to be equal to that of the star and this gives a value of $10^{12}$ K . However, since the enhancement happens for a small range of rotational phases, the actual emitting region must be much smaller than what has been assumed. Thus the above estimation can only be treated as a lower limit to the actual brightness temperature. The nearly 100 per cent circular polarization,  together with the order-of- magnitude enhancement, and high brightness temperature,  indicate a coherent emission process. 
As mentioned in \S \ref{sec:intro}, both ECME and plasma  mechanisms can plausibly give rise to this kind of phenomenon. 
However, in case of ECME, energy is radiated away at an angle given by $\theta=\cos^{-1}(v/c)$  ($v$ is the characteristic speed of the electron population at resonance with the electromagnetic wave and $c$ correspond to the  and the speed 
 of light) with respect to the direction of
 the magnetic field \citep{melrose82}.  For an electron distribution typical of a magnetic hot star, $v$ is such that $\theta \sim 90^\circ$ \citep{trigilio08}.
This means that the flux enhancement should be obtained when the line of sight magnetic field is close to zero. This is exactly the case we observed (see Figure \ref{fig:f1}). 
Since plasma emission has  no such dependence on magnetic field in terms of  direction of emission, we conclude that ECME is greatly favoured over plasma emission as the source of the observed coherent emission.

\citet{lim96,bailey12} first reported that the 5 and 8\,GHz flux density of HD\,133880 is modulated with the stellar rotation period, with stronger, broader peaks near the
 longitudinal field extrema (phases 0.0 and 0.5) and weaker, narrower peaks near the longitudinal field nulls
 (phases 0.25 and 0.75). 
   They also found that the degree and sense of circular polarization 
were smoothly modulated, with maximum circular polarization reaching $\sim\pm16$ per cent at phase $\sim 0.5$ with a mean of 0,  and sense of circular polarization corresponding to the 
extraordinary mode of propagation. 
 Thus the ECME mechanism, more favorable at low frequencies, appears to be triggered somewhere between
0.6--5.0\,GHz. In CU Vir, until now the only star for which ECME has been confirmed, the emission at 2\,GHz was found to be ECME \citep{ravi10}. More studies in this frequency range  are needed to ascertain the nature of the contributing emission mechanisms as a function of frequency.

\begin{figure}
\centering
\includegraphics*[width=0.4\textwidth]{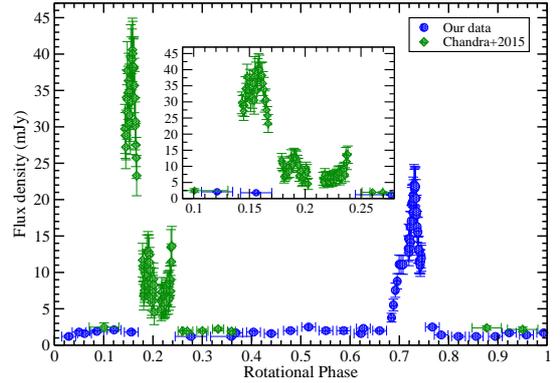}
\caption{Variation of the 610 MHz flux of HD 133880 with rotational phase for $RR^*$ polarization, i.e. Stokes $(I+V)$. Here we have combined our data with those 
of \citet{chandra15}.  The inset is the zoomed version of the phase range from 0.1--0.28, covering the flux enhancement reported by \citet{chandra15}, which  is not detected in our data which covers a fraction of this phase range.
 \label{fig:f2}}
\end{figure}

\citet{george12,chandra15}  also reported the $610$ MHz GMRT observations of HD\,133880. However,  the observations were of short duration 
and did not cover the phase 0.731 at which we observed the flux enhancement. 
However, it is important to note that \citet{chandra15} reported a  large flux enhancement at 1390 and 610\,MHz  at phases $0.827\pm0.012$ and $0.261\pm0.030$,
respectively  \cite[using the ephemeris of][]{bailey12}. 
While they indicated the possibility of ECME,  they did not have full coverage or  polarization information to derive any conclusions, as their data were taken in dual band mode \citep[610/235 MHz,]{george12}, and as a consequence they measured only $RR^*$ polarization at 610 MHz. One interesting thing to note here is that their 1390 MHz phase of enhancement now becomes (after using our ephemeris) $0.725\pm 0.012$, thus matching the 610 MHz phase of enhancement of our data. This increases the necessity to make a similar study at L band since the previous study by \cite{chandra15} suffers from sparse phase coverage and lack of polarisation information.

With the ephemeris adopted here, the flux enhancement observed by \citet{chandra15} corresponds to a phase of $0.172\pm0.030$ - close to one-half rotation away from the enhancement we detect, and approximately coincident with the other null of the longitudinal field. 
We reanalysed these data with better  time resolution (1 minute) during the phase of enhanced emission and derived the phase of maximum flux enhancement to be $0.159\pm0.001$.
Our new data cover this
phase, but we do not see a flux enhancement (Fig. \ref{fig:f2}).  

There could be two plausible explanations for this.  
First, the previous event may have been a transient phenomenon. \citet{ravi10} reported significant variations in pulse height  of  ECME radio flux enhancements of CU\,Vir 
 on a timescale of $\sim 200$ rotations. In fact, at 13\,cm, one of the pulses  disappeared completely. The time separation of the 610 MHz observations obtained by us and by \citet{chandra15} corresponds to 2254 days, or about 2569 rotations of HD\,133880. Hence it is plausible that a similar evolution of the magnetosphere of HD\,133880 has occurred.
Alternatively, the peak could still be present, but may fall into the small phase window ($0.170-0.245$) in which we lack data. Shifts in the phases of the radio emission peaks were also noted in the case of CU Vir.  However, \citet{ravi10} could not establish if the phase shifts were due to intrinsic variations of the stellar rotation period \citep[long known to occur for CU\,Vir; e.g.][]{pyper13} or instabilities in the region of emission \citep{ravi10}. 
If the emission region is unstable (or drifting) for HD\,133880, then it would need to have drifted in phase by at least 0.01 cycles since December, 2009 for us to have missed it. Alternatively, if the phase shift is a consequence of a period change, the period of the star would need to have changed by approximately 0.34 second per day since 2009.
To put this in context, \cite{kochukhov17} reported evidence exists for a 0.6 second net increase in the rotational period of HD\,133880 between 1974--1988 and 2010--2015. While at face value the required period change seems rather large, the short-term behaviour of the period evolution is entirely unknown.

\par 
In Figure \ref{fig:f2}, we have combined our $RR^*$ polarization data  with those of \cite{chandra15}. The enhancement around phase 0.16 \citep[data from][]{chandra15} is clearly much larger than that around phase 0.73. A similar phenomenon was also observed for CU\,Vir \citep{trigilio11}, which was attributed to instability in the emission region \cite{ravi10}. However, it is interesting to note that in both  CU Vir and HD\,133880, the stronger pulse was obtained when the North magnetic pole was approaching the observer and weaker pulse was obtained when the North pole was receding away. Thus, the difference in ECME pulse height might be a result of a more fundamental phenomenon and this certainly deserves more attention.

\par 
Another important result is the absence of enhancement in left circular polarization, also seen in  CU\,Vir. \citet{trigilio00} attributed this to the deviation of the magnetic field from a perfect dipole. 
The extensive study of the magnetic field of HD\,133880 by \cite{kochukhov17} has revealed that the star has an extended but weaker positive field region and a smaller but stronger negative field region. For such a configuration it may well be possible to distort the magnetosphere in such a way that the coherent emission from the South pole experiences much stronger absorption in the inner magnetosphere than that from the North pole.

It is interesting to note that CU\,Vir and HD\,133880 are remarkably similar. Both are rapidly-rotating late-B stars with distorted dipole fields. Both show evidence for period changes, and  exhibit radio pulses attributable to ECME. 
As illustrated by these results, the detection of periodic pulses can act as a sensitive probe of any change in the stellar rotation period, since in the case of a period change, the phases of arrival of the pulses will vary with time  \citep{trigilio08}. Thus a detailed study of the light curve of HD\,133880 at multiple radio frequencies might help in resolving the period issue (Das et al., in preparation).
 
To summarise, we present the rotational variation of the 610 MHz flux of the Bp star HD\,133880. We observe order of magnitude flux enhancement  at phase $0.731\pm 0.001$ in $RR^*$ polarization. No flux enhancement is obtained for $LL^*$ polarization at any phase, which we attribute to the fact that the star has a magnetic field which is significantly different from a pure dipole \citep{landstreet90,bailey12,kochukhov17}. Considering the high brightness temperature, $\sim 100$ per cent circular polarization and the almost null value of the line of sight magnetic field $B_z$ at the peak of emission, we infer that the emission mechanism is likely to be coherent radio mechanism due to ECME.  Further studies to understand the complete morphology and characteristics of the radio emission are underway.

\vspace{-0.3cm}
\section*{Acknowledgements}
We thank the referee for very useful comments which helped us in improving the manuscript.
PC  acknowledges support from the Department of Science and Technology via 
SwarnaJayanti Fellowship awards (DST/SJF/PSA-01/2014-15).  We thank the staff of the GMRT that made these observations possible. 
The GMRT is run by the National Centre for Radio Astrophysics of the Tata Institute 
of Fundamental Research. GAW acknowledges Discovery Grant support from the Natural Sciences and Engineering Research Council (NSERC) of Canada.









\bsp	
\label{lastpage}
\end{document}